\documentclass[11pt]{article}
\usepackage{amssymb}
\usepackage{amsthm}
\usepackage{amsmath}
\usepackage{amsbsy}
\usepackage{epsfig}
\usepackage{color}
\usepackage{enumerate}
\usepackage{vmargin}
\setpapersize{USletter}
\setmarginsrb{1in}{1in}{1in}{1in}{0pt}{0mm}{0pt}{36pt}

\begin{document}

\title{{\bf\Large Intrinsic Random Functions and Universal Kriging on the Circle}}
\author{Chunfeng Huang$^1$, Haimeng Zhang$^2$, and Scott M. Robeson$^3$}
 \maketitle

\date{}

{\bf Abstract.} Intrinsic random functions (IRF) provide a versatile approach when the assumption of second-order stationarity is not met. Here, we develop the IRF theory on the circle with its universal kriging application.  Unlike IRF in Euclidean spaces, where differential operations are used to achieve stationarity, our result shows that low-frequency truncation of the Fourier series representation of the IRF is required for such processes on the circle. All of these features and developments are presented through the theory of reproducing kernel Hilbert space. In addition, the connection between kriging and splines is also established, demonstrating their equivalence on the circle.

\footnotetext[1]{Departments of Statistics and Geography, Indiana University,
Bloomington, IN 47408 U.S.A.; huang48@indiana.edu.}

\footnotetext[2]{Department of Mathematics and Statistics,
University of North Carolina at Greensboro, Greensboro, NC 27402 U.S.A.; Phone  +1 (336) 334-5836; Fax +1 (336) 334-5949; haimengzhanguncg@gmail.com; Corresponding Author.}

\footnotetext[3]{Departments of Geography and Statistics, Indiana
University, Bloomington, IN 47405 U.S.A.;
srobeson@indiana.edu}

\newpage

{\bf 1. Introduction.} When a random process is considered on a circle, it is often assumed to be second-order stationary (or stationary for short in the paper), that is, the mean of the process is constant over the circle and the covariance function at any two points depends only on their angular distance (Yaglom, 1961, Roy, 1972, Roy and Dufour, 1974, Dufour and Roy, 1976, Wood, 1995, Gneiting, 1998). While stationarity is commonly assumed, it is often considered to be unrealistic in practice. In Euclidean spaces, a large class of non-stationary phenomena may be represented through intrinsic random functions (IRF, Matheron, 1973, Cressie, 1993, Chil\`es and Delfiner, 2012). The properties of IRF in other spaces, such as the circle or sphere, are not widely known. In this paper, the theory of IRF on the circle is developed, where we find that instead of differential operations, truncation of the Fourier series representation becomes essential for IRF on the circle. This can be presented in the context of the reproducing kernel Hilbert space (RKHS, Aronszajn, 1950, Wahba, 1990a). We formally make such a connection and further relate universal kriging with the smoothing formula in RKHS. Based on this approach, we are able to demonstrate the equivalence between splines and kriging on the circle. \\

{\bf 2. IRF and RKHS.} A key component for IRF is the allowable measure. Based on Matheron (1973) and Chil\`es and Delfiner (2012, Chapter 4), a discrete measure $\lambda = \sum_{i=1}^m \lambda_i \delta(t_i)$ on a unit circle $S$, where $t_i \in S, \lambda_i \in\mathbb{R}$ and $\delta(\cdot)$ is the Dirac measure, is allowable at the order of an integer $\kappa (\kappa \ge 0)$ if it annihilates all trigonometric functions of order $k < \kappa$. That is,
\begin{equation} \label{eq:lambdakappa}
\sum_{i=1}^m \lambda_i \cos (k t_i) = \sum_{i=1}^m \lambda_i \sin (k t_i) = 0, \quad 0 \le k < \kappa.
\end{equation}
We call $\Lambda_{\kappa}$ the class of such allowable measures. Clearly $\Lambda_{\kappa+1} \subset \Lambda_{\kappa}$. In addition, for $\lambda \in \Lambda_\kappa$, the translated measure $\iota_t \lambda = \sum_{i=1}^m \lambda_i \delta(t_i + t), t \in S$ remains in $\Lambda_{\kappa}$. This can be easily seen from the elementary trigonometric identities (also see Matheron, 1979, Chil\`es and Delfiner, 2012).  For any function $f(\cdot)$ on $S$, we define $f(\lambda) = \sum_{i=1}^m \lambda_i f(t_i)$. \\

In this paper, we consider a random process $\{ Z(t), t \in S\}$ on a unit circle with finite second moment and being continuous in quadratic mean. By Yalgom (1961) and Roy (1972), the process can be expanded in a Fourier series which is convergent in quadratic mean:
\begin{equation} \label{eq:quadexpansion}
Z(t) = Z_0 + \sum_{n=1}^\infty (Z_{n,c} \cos n t + Z_{n,s} \sin n t),
\end{equation}
where $Z_0 = 1/(2\pi) \int_S Z(t) dt, Z_{n,c} = (1/\pi)\int_S Z(t) \cos n t dt,$ and $Z_{n,s} =  (1/\pi) \int_S Z(t)\sin n t dt.$ \\

\noindent
{\bf Definition 2.1}. For an integer $\kappa (\kappa \ge 0)$, the random process in (\ref{eq:quadexpansion}) is called an IRF$\kappa$ if for any $\lambda \in \Lambda_{\kappa}$, the process
\[
Z_{\lambda}(t) = Z(\iota_t \lambda) = \sum_{i=1}^m \lambda_i Z(t_i + t)
\]
is stationary with respect to $t \in S$ and has a zero mean. \\

To characterize such a circular IRF$\kappa$, we denote
\[
Z_{\kappa}(t) = \sum_{n=\kappa}^\infty (Z_{n,c} \cos n t + Z_{n,s} \sin n t),
\]
as its low-frequency truncated process and so we have the following Lemma. \\

\noindent
{\bf Lemma 2.1.} A random process given by (\ref{eq:quadexpansion}) is an IRF$\kappa$ if and only if its low-frequency truncated process $Z_{\kappa}(t)$ is stationary and has a zero mean.

\begin{proof} In the Fourier expansion (\ref{eq:quadexpansion}), the lower trigonometric functions will be annihilated by $\lambda \in \Lambda_{\kappa}$, which implies
\[
Z(\iota_t \lambda) = \sum_{n=\kappa}^\infty \left( Z_{n,c} \sum_{i=1}^m (\lambda_i \cos n t_i \cos n t - \lambda_i \sin n t_i \sin n t) + Z_{n,s} \sum_{i=1}^m \lambda_i \sin n t_i \cos n t + \cos n t_i \sin n t) \right).
\]
Denote $\lambda_{n,c} = \sum_{i=1}^m \lambda_i \cos n t_i, \lambda_{n,s} = \sum_{i=1}^m \lambda_i \sin n t_i$ and
\begin{equation} \label{eq:YZ}
Y_{n,c} = Z_{n,c} \lambda_{n,c} + Z_{n,s} \lambda_{n,s}, \quad Y_{n,s} = - Z_{n,c} \lambda_{n,s} + Z_{n,s} \lambda_{n,c}, \quad n=\kappa, \kappa+1,\ldots,
\end{equation}
we have
\[
Z(\iota_t \lambda) = \sum_{n=\kappa}^\infty (Y_{n,c} \cos n t + Y_{n,s} \sin n t).
\]
Yaglom (1961, Theorem 5) shows that a random process (\ref{eq:quadexpansion})  on the circle is stationary if and only if its Fourier coefficients are uncorrelated random variables.   The Lemma can be directly obtained based on this and the linear mapping between the coefficients of $(Z_{n,c}, Z_{n,s})$ and $(Y_{n,c}, Y_{n,s})$ in (\ref{eq:YZ}).
\end{proof}

{\bf Remark 2.1.} In Euclidean spaces, the IRF is associated with differential operations (Matheron, 1973, Chil\`es and Delfiner, 2012). For example, a differentiable IRF$\kappa$ on a real line is characterized as that its $(\kappa+1)$ derivative is stationary. Lemma 2.1 indicates that for circular processes, the low-frequency truncation operation replaces differential operations and leads to stationarity. This observation also has important implications for splines on the circle, which is addressed in Section 4. \\

{\bf Remark 2.2.} It is clear that an IRF$0$ on the circle is the conventional stationary process. Note that this is slightly different from what has been defined in the Euclidean spaces, where IRF$(-1)$ is usually a stationary process. For the rest of this paper, we assume $\kappa \ge 1$ for notational simplicity. \\

{\bf Remark 2.3.} Based on Lemma 2.1, for an IRF$\kappa$ process $Z(t)$, the random process
\[
Z^*(t) = Z(t) + A_0 + \sum_{n=1}^{\kappa-1} (A_{n,c} \cos n t + A_{n,s} \sin nt),
\]
where $A_0, A_{n,c}, A_{n,s}, n=1, \ldots, (\kappa-1)$ are random variables, is clearly also an IRF$\kappa$. These two processes $Z(t)$ and $Z^*(t)$ share the same truncation process $Z_{\kappa}(t)$, with $Z^*(\lambda) = Z(\lambda)$, for any $\lambda \in \Lambda_{\kappa}$. Similar to the discussion in Chil\`es and Delfiner (2012, Section 4.4.2), these functions form an equivalent class. \\

To obtain the covariance function of $Z_{\kappa}(t)$ of an IRF$\kappa$, based on Yaglom (1961) or Roy (1972), we denote
\[
\mbox{E} (Z_{n,c} Z_{m,c}) = \mbox{E}(Z_{n,s} Z_{m,s}) = \delta(n-m) \gamma_n,
\]
with $\gamma_n > 0, n \ge \kappa$, and $\sum_{n=\kappa}^\infty \gamma_n < \infty$. Here $\delta(n-m) = 1$ if $n=m$ and 0 otherwise. This leads to
\[
\mbox{cov} (Z_{\kappa}(x), Z_{\kappa}(y)) = \sum_{n=\kappa}^\infty \gamma_n \cos n (x-y) : = \phi(x-y), \quad \quad x, y \in S.
\]

\noindent This covariance function $\phi(\cdot)$ plays an essential role in our paper, and is named as the {\it intrinsic covariance function} of the IRF$\kappa$.\\

{\bf Remark 2.4.} Another component in IRF is the generalized covariance function. For an IRF$\kappa$ on the circle, it is clear that $\phi(\cdot)$ is a candidate of te generalized covariance function. By the annihilation property of allowable measures, any function which is the sum of $\phi(\cdot)$ and a linear combination of lower trigonometric functions up to order $\kappa-1$ can also be a generalized covariance function. A basic property of the generalized covariance function in Euclidean spaces is its conditional positive definiteness.  Such a property (Levesley et al. 1999) seems less important on the circle.  One can clearly show that $\phi(\cdot)$ is conditional positive definite of order $\kappa$, and positive definite simultaneously (Schoenberg, 1942). For example, it is noted by Yaglom (1961) that the space of valid variograms coincides with the space of valid covariances.  \\

RKHS was introduced in Aronszajn (1950), and is popularly used in the spline literature (see, for example, Wahba, 1990a). Taijeron et al. (1994) and Levesley et al. (1999) have studied RKHS in the context of spline interpolation and smoothing on the circle and sphere. Here, we formally establish the connection between IRF and RKHS on the circle. First, given an IRF$\kappa$ process and its intrinsic covariance function $\phi(\theta) = \sum_{n=\kappa}^\infty \gamma_n \cos n \theta, \gamma_n > 0$, one can define a function space $X_\kappa$ on $S$, following Levesley et al. (1999)
\[
X_{\kappa} = \left\{ f(t) = a_0 + \sum_{n=1}^\infty (a_{n,c} \cos nt + a_{n,s} \sin nt), t \in S: \quad \sum_{n=\kappa}^\infty \frac{1}{\gamma_n} (a_{n,c}^2 + a_{n,s}^2) < \infty \right\}.
\]
For $f, g \in X_{\kappa}$, a semi-inner product is defined
\begin{equation} \label{eq:seminorm}
\langle f, g \rangle_{\kappa} = \sum_{n=\kappa}^\infty \frac{1}{\gamma_n} (a_{n,c,f} a_{n,c,g} + a_{n,s,f} a_{n,s,g}).
\end{equation}
There is a nil space for this semi-inner product $N=\mbox{span} \{ 1, \cos t, \sin t, \ldots, \cos (\kappa-1) t, \sin (\kappa-1) t \}$. Denote $l = \mbox{dim} (N)$, and it is clear that $l = 2 \kappa-1$. Let $\{ \tau_1, \ldots, \tau_l \in S \}$ be a set of distinct points such that for every $p(\cdot) \in N$, if $p(\tau_{\nu})=0$ for all $\nu=1,\ldots, l$, then $p \equiv 0$. Then, the inner product
\[
\langle f, g \rangle = \sum_{\nu=1}^l f(\tau_{\nu}) g(\tau_{\nu}) + \langle f, g \rangle_{\kappa}
\]
is well defined and $X_{\kappa}$ can be shown to be complete with respect to the norm induced by this inner product (Levesley et al. 1999). In addition, there exist $p_1(t), \ldots, p_l(t) \in N$, such that $p_{\nu}(\tau_{\mu}) = \delta(\nu,\mu)$ for $ 1 \le \nu, \mu \le l$. As discussed in Levesley et al. (1999), the space $X_{\kappa}$ is a Hilbert function space in which point evaluations are continuous linear functionals. Therefore, for $x, y \in S$, there exists a reproducing kernel
\begin{equation} \label{eq:Hxy}
H(x,y) = \phi(x-y) - \sum_{\nu=1}^l \left(\phi(x-\tau_{\nu}) p_{\nu}(y) + \phi(y-\tau_{\nu}) p_{\nu}(x)\right) + \sum_{\nu=1}^l \sum_{\mu=1}^l \phi(\tau_{\nu}-\tau_{\mu}) p_{\nu}(x) p_{\mu}(y)+ \sum_{\nu=1}^l p_{\nu}(x) p_{\nu}(y).
\end{equation}

\noindent
{\bf Proposition 2.1.} $H(x,y)$ is positive definite for $x, y \in S$.

\begin{proof} By the expansion of $\phi(\theta) = \sum_{n=\kappa}^\infty \gamma_n \cos (n \theta)$, we have
\begin{eqnarray*}
&\,& H(x,y) = \sum_{\nu=1}^l p_{\nu}(x) p_{\nu}(y) + \sum_{n=\kappa}^\infty \gamma_n \left( \cos (n x) - \sum_{\nu=1}^l \cos (n \tau_{\nu}) p_{\nu}(x) \right) \left( \cos (n y) - \sum_{\nu=1}^l \cos (n \tau_{\nu}) p_{\nu}(y) \right) \\
&\,& \quad + \sum_{n=\kappa}^\infty  \gamma_n \left( \sin (n x) - \sum_{\nu=1}^l \sin (n \tau_{\nu}) p_{\nu}(x) \right) \left( \sin (n y) - \sum_{\nu=1}^l \sin (n \tau_{\nu}) p_{\nu}(y) \right).
\end{eqnarray*}
The positive definiteness of $H(x,y)$ can be obtained through the symmetry in $x$ and $y$ along with $\gamma_n > 0$ for all $n \ge \kappa$.
\end{proof}

{\bf Remark 2.5.} We show that for an IRF$\kappa$ on the circle, there exists a corresponding RKHS. Conversely, given the positive definiteness of $H(x,y)$, there exists a Gaussian random process that is an IRF$\kappa$ with $H(x,y)$ as its covariance function. Therefore, the connection between IRF and RKHS on the circle has been formally established.\\

{\bf Remark 2.6.} One can easily verify the reproducing property of this kernel (Taijeron et al. 1994, Light and Wayne, 1995, Levesley et al. 1999). The different choice of the distinct points $\{\tau_1, \ldots, \tau_l\}$ will alter the form of the reproducing kernel with $\phi(\cdot)$ unchanged since the terms containing $p_{\nu}(\cdot)$ will all be annihilated by $\lambda \in \Lambda_{\kappa}$.   \\

{\bf 3. Univesal kriging.} Universal kriging associated with IRF is widely used in spatial statistics (Cressie, 1989, Chil\`es and Delfiner, 2012). In this Section, we will discuss universal kriging on the circle. Let $Z(t)$ be an IRF$\kappa$ with an intrinsic covariance function $\phi(\cdot)$ and the mean function $\mbox{E} (Z(t)) = \sum_{\nu=1}^l \beta_{\nu} q_{\nu}(t)$, where $\beta_{\nu}, \nu=1,\ldots, l$ are coefficients  and $\mbox{span}\{ q_1(t), \ldots, q_l(t) \} = N$. Here $q_{\nu}(\cdot)$ can be the elementary lower trigonometric functions or $\{p_{\nu}(t)\}_{\nu=1}^l$ given in Section 2. Assume that the data $\{ (t_i, y_i), i=1\ldots, n\}, n \ge l$, are observed from this IRF$\kappa$ with measurement error
\[
Y(t) = Z(t) + \epsilon(t), \quad t \in S,
\]
where $\epsilon(\cdot)$ is a white noise process with mean zero that is uncorrelated with the process $Z(\cdot)$.  \\

To obtain the best linear unbiased estimator at $t_0 \in S$, the universal kriging is commonly used, where the linear estimator is
\[
\hat{Z} (t_0) = \eta^T y, \quad y=(y_1, \ldots, y_n)^T,
\]
with coefficients $\eta = (\eta_i)_{n \times 1}$.  The unbiasedness leads to $\eta^T Q = q^T,$ where
\[
\quad Q = \{ q_{\nu}(t_i)\}_{n \times l}, \quad q= (q_1(t_0), \ldots, q_l (t_0))^T,
\]
and therefore $\sum_{i = 1}^n \eta_i q_j(t_i) = q_j(t_0), j = 1, 2 , \ldots, l$, which implies
\begin{eqnarray} \label{unbiasedness}
\sum_{i=1}^n \eta_i \delta(t_i) - \delta(t_0) \in \Lambda_{\kappa}.
\end{eqnarray}
Hence, the squared prediction error can be shown to be
\[
\mbox{E} (\hat{Z}(t_0) - Z(t_0))^2 = \sigma^2 \eta^T \eta+ \eta^T \Psi \eta - 2 \eta^T \phi + \phi(0),
\]
where
\[
\Psi = \{ \phi(t_i - t_j) \}_{n \times n}, \quad \phi = (\phi(t_1-t_0), \ldots, \phi(t_n-t_0))^T.
\]
The goal of universal kriging is to minimize the squared prediction error, subject to the unbiasedness constraints. Letting a vector $\rho$ of  $l\times 1$ be the Lagrange multipliers, we need to minimize
\[
M(\eta) = \sigma^2 \eta^T \eta + \eta^T \Psi \eta - 2 \eta^T \phi + \phi(0) + 2  (\eta^T Q - q^T) \rho.
\]
Direct computation finds the universal kriging formula as
\begin{equation} \label{eq:uk1}
\left\{ \begin{array}{l} (\Psi + \sigma^2 I) \eta + Q \rho = \phi, \\ Q^T \eta = q. \end{array} \right.
\end{equation}


Next, we show that this universal kriging formula can be interpreted in the content of RKHS. With this IRF$\kappa$ $Z(\cdot)$ and the observed data $\{ (t_i, y_i), i=1\ldots, n\}, n \ge l$, the smoothing problem is to find a function $f_{\alpha}(t) \in X_{\kappa}$ such that it minimizes (Taijeron et al. 1994, Levesley et al. 1999)
\[
\sum_{i=1}^n (y_i - f(t_i))^2 + \alpha \| f \|_{\kappa}^2,
\]
where $\alpha > 0$ is the smoothing parameter and $\| \cdot \|_{\kappa}$ is induced by the semi-inner product (\ref{eq:seminorm}) in Section 2. The minimizer can be shown to be
\begin{equation} \label{eq:smoothing}
f_{\alpha}(t) = \sum_{\nu=1}^l d_{\nu} q_{\nu}(t) + \sum_{i=1}^n c_i \phi(t_i - t),
\end{equation}
where $c=(c_i)_{n \times 1}$ and $d = (d_{\nu})_{l \times 1}$ satisfy the following,
\[
\left\{ \begin{array}{l} (\Psi + \alpha I) c + Q d = y, \\ Q^T c = 0_{l \times 1}. \end{array} \right.
\]
To show the connection between this smoothing formula and universal kriging (\ref{eq:uk1}), note that the smoothing formula for an unobserved point $t_0 \in S$ can be rewritten in the following manner
\begin{eqnarray*}
&\,& f_{\lambda}(t_0) = (c,d)^T \left( \begin{array}{l} \phi \\ q \end{array} \right)  = (y^T, 0_{1 \times l}) \left( \begin{array}{cc} \Psi + \alpha I & Q \\ Q^T & 0_{l \times l} \end{array} \right)^{-1} \left( \begin{array}{c} \phi \\ q \end{array} \right) \\
&\,& \quad := (y^T, 0_{1 \times l}) \left( \begin{array}{c} \eta^* \\ \rho^* \end{array} \right) = \eta^{*T} y,
\end{eqnarray*}
where
\[
\left( \begin{array}{c} \eta^* \\ \rho^* \end{array} \right) = \left( \begin{array}{cc} \Psi + \alpha I & Q \\ Q^T & 0_{l \times l} \end{array} \right)^{-1} \left( \begin{array}{c} \phi \\ q \end{array} \right),
\]
or
\begin{equation} \label{eq:uk2}
\left\{ \begin{array}{ll} (\Psi + \alpha I) \eta^* + Q \rho^* = \phi, \\ Q^T \eta^* = q. \end{array} \right. \\
\end{equation}

{\bf Remark 3.1.} This equation (\ref{eq:uk2}) is exactly the dual formula of universal kriging (Cressie, 1993, Chil\`es and Delfiner, 2012). Usually, universal kriging is viewed as a linear estimator of observed data and the smoothing formula is viewed as linear combination of the intrinsic covariance with lower trigonometric trends. From the above discussion, these two views are essentially the same. The connection between universal kriging and the smoothing formula, therefore, is obvious.\\

{\bf Remark 3.2.} The smoothing parameter $\alpha$ in (\ref{eq:uk2}) and the noise variance $\sigma^2$ in (\ref{eq:uk1}) play the same role. For example, in the smoothing formula, when $\alpha$ increases to infinity, the minimization procedure demands $\| f \|_{\kappa} \downarrow 0$, which shows that $c \to 0$, and the smoothing formula reduces to the trigonometric regression (Eubank, 1988). In kriging practice, when $\sigma^2$ increases to infinity, the noise overwhelms the spatial dependency, the process becomes uncorrelated. The squared prediction error is dominated by $\sigma^2 \eta^T \eta$. The universal kriging reduces to minimize $\eta^T \eta$ subject to the unbiasedness restriction, which also leads to trigonometric regression prediction. When both $\alpha$ and $\sigma^2$ decrease to zero, both smoothing and kriging result in exact interpolation. \\

{\bf Remark 3.3.} In this paper, we extend IRF to the circular setting through the RKHS theory. In so doing, we find that the lower monomials in Euclidean spaces need to be replaced by lower trigonometric functions, and the differential operators need to be replaced by low-frequency truncations. The RKHS sheds light into kriging on the circle and provides elementary understanding. This RKHS approach allows us to revisit splines on the circle, see Section 4.1.  \\

{\bf 4. Examples and discussions.}

{\bf 4.1 Spline on the circle.} In Craven and Wahba (1979) and Wahba (1990a), the spline on the circle is the minimizer of
\[
\sum_{i=1}^n (y_i - f(t_i))^2 + \alpha J(f),
\]
where
\begin{equation} \label{eq:penalty}
J(f) = \int_S (f^{(m)}(t))^2 d t,
\end{equation}
This minimizer has been shown to be (Wahba, 1990a)
\[
f_{\alpha}(t) = d + \sum_{i=1}^n c_i R(t_i, t),
\]
with
\[
R(s,t) = 2\sum_{n=1}^\infty \frac{1}{n^{2m}} \cos n (s-t).
\]

{\bf Remark 4.1}. This is the smoothing formula (\ref{eq:smoothing}) for $\kappa=1$, where the intrinsic covariance function $\phi(\cdot)$ is replaced by $R(s,t)$. Note that the spline kernel $R(s,t)$ is a specified function, for example,
\begin{eqnarray*}
&\,& m=1, \quad R(s,t) = \frac{(s-t)^2}{2} - \pi|s-t| + \frac{\pi^2}{3}, \\
&\,& m=2, \quad R(s,t) = -\frac{(s-t)^4}{24} + \frac{\pi |s-t|^3}{6} - \frac{\pi^2 (s-t)^2}{6} + \frac{\pi^4}{45}.
\end{eqnarray*}
These kernels $R(s,t)$ are clearly positive definite (Schoenberg, 1942) on the circle, and are valid covariance functions.  \\

{\bf Remark 4.2}. In Euclidean spaces, the order $m$ in (\ref{eq:penalty}) plays a significant role. It indicates the smoothness assumption of the function, and relates to the order of IRF in kriging (Kent and Mardia, 1994). However, for this spline on the circle, this order $m$ only alters the covariance functions (see Remark 4.1), and loses its connection to the order $\kappa$ of IRF. Therefore, the spline with derivative penalty has limited application for circular processes. As shown in Lemma 2.1, the low-frequency truncation operation shall be used, leading to the more appropriate spline model on the circle
\begin{eqnarray} \label{spline-model}
\frac{1}{n} \sum_{i=1}^n (y_i - f(t_i))^2 + \alpha \| f \|_{\kappa}^2,
\end{eqnarray}
where $\| \cdot \|_{\kappa}$ is induced by the semi-inner product (\ref{eq:seminorm}) in Section 2. A more general approach for hyperspheres can be found in Taijeron et al. (1994). \\


{\bf 4.2. Splines and kriging.} The spline model (\ref{spline-model}) is exactly the same as the smoothing formula in Section 3, where the equivalence between the smoothing formula and kriging is discussed. It is clear that this spline (\ref{spline-model}) and kriging are also equivalent on the circle. The RKHS theory on the circle offers a clear view of this connection. In addition, following Remark 3.1, spline is a linear combination of the intrinsic covariance functions with lower trigonometric trends, while universal kriging is a linear estimator of observed data. They arrive at the same conclusion as dual formulations of kriging (Cressie, 1993).\\

The connections between splines and kriging have been extensively discussed in literature, including Matheron (1981), Watson (1984), Lorenc (1986), Cressie (1989, 1990, 1993), Wahba (1990a, 1990b), Kent and Mardia (1994), Laslett (1994), Furrer and Nychka (2007) among others. Furrer and Nychka (2007) show that, given a covariance function, one can construct a reproducing kernel and obtain a general spline estimate in Euclidean space. In this paper, we show this connection formally for circular processes using the IRFs.\\

{\bf 4.3. Ordinary kriging.} The ordinary kriging is well known and has been widely used in a variety of spatial analysis contexts. On the circle, ordinary kriging is equivalent to the universal kriging developed in Section 3 with $\kappa=1$. The process $Z(t)$ is IRF$1$ with mean and covariance
\[
\mbox{E}(Z(t)) = \beta_1, \quad \mbox{cov} (Z(x), Z(y)) =H(x,y) =  \phi(x-y) - \phi(x-\tau_1) - \phi(y-\tau_1) + \phi(0) + 1.
\]
The process is not stationary, but a direct computation gives
$\mbox{var} (Z(x)-Z(y)) = 2\phi(0) - 2 \phi(x-y).$
That is, the variance of the process at two points only depends on their circular distance and we can define
\[
\tau(\theta) := \phi(0)-\phi(\theta) = \sum_{n=1}^\infty \gamma_n (1-\cos n \theta),
\]
as the semi-variogram (Schoenberg 1942, Huang et al. 2011). The process can be viewed as intrinsically stationary on the circle. Note that $\phi(\theta)$ relates to the semi-variogram directly through (Huang et al. 2011)
\[
\phi(\theta) = c_0 - \tau(\theta), \quad c_0 \ge \frac{1}{\pi} \int_{0}^{\pi} \tau(\theta) d \theta.
\]
Therefore, with  $\tau=(\tau(t_1-t_0), \ldots, \tau(t_n-t_0))^T, 1_n = (1, 1, \cdots, 1)^T$, and  $\Gamma=\{ \tau(t_i-t_j)\}$, we have the kriging estimator
\[
f_{\alpha}(t_0) = \eta^T y,
\]
where $\eta$ satisfies
\[
-\Gamma \eta + \rho 1_n = \tau, \quad 1_n^T \eta =1.
\]
This is exactly the ordinary kriging in the spatial literature (Cressie, 1993). \\

{\bf 4.4. Brownian bridge.} The Brownian bridge $\{ B(t), t \in S \}$ is a random process on the circle with mean zero and
\[
\mbox{cov} (B(s), B(t)) = 2 \pi \min \{ s,t \} - st, \quad s, t \in S.
\]
It is clearly not a stationary process. However, for any allowable measure $\lambda = \sum_{i=1}^m \lambda_i \delta(t_i) \in \Lambda_1$ (hence $\sum_{i=1}^m \lambda_i = 0$),
\begin{eqnarray*}
&\,& \mbox{cov} (B(\iota_t \lambda), B(\iota_s \lambda)) = \sum_{i,j=1}^m \lambda_i \lambda_j \{2 \pi \min \{ t_i + t, t_j + s \} - (t_i +t)(t_j+s) \} \\
&\,& = - \pi \sum_{i,j=1}^m \lambda_i \lambda_j |t_i+t-t_j - s| + (1/2) \sum_{i,j=1}^m \lambda_i \lambda_j (t_i + t-t_j - s)^2,
\end{eqnarray*}
where the last equality is based on $2\min \{ a,b\}= (a+b)-|a-b|, a > 0, b > 0$ and $\sum_{i=1}^m \lambda_i = 0$. That is, $B(\iota_t \lambda)$ is stationary with respect to $t$, showing that the Brownian bridge is an IRF$1$. \\

On the other hand, one can consider the Fourier series representation of the Brownian bridge
\[
B(t) = B_0 + \sum_{n=1}^\infty ( B_{n,c} \cos (nt) + B_{n,s} \sin (nt)).
\]
Direct computation shows that for all $n,m \ge 1, \mbox{E}(B_{n,c} B_{m,c}) = \mbox{E}(B_{n,s} B_{m,s}) = \frac{2}{n^2}\delta(m,n)$ and $\mbox{E}(B_{n,c}B_{m,s}) = 0$. Therefore, the truncated process $B_1(t) = \sum_{n=1}^\infty (B_{n,c} \cos (nt) + B_{n,s} \sin (nt))$ is clearly stationary, and so the Brownian bridge is an IRF$1$ by Lemma 2.1. In addition, its intrinsic covariance function is given by $\phi(s-t) = 2\sum_{n=1}^\infty \frac{1}{n^2}\cos(n(s-t))$, which is exactly the same as the $R(s, t)$ with $m = 1$ in Subsection 4.1. However, we have $\mbox{E}(B_0 B_{n,c}) = -\frac{2}{n^2}, n \ge 1$, that is, $B_0$ is correlated with all $B_{n,c}, n \ge 1$. Such a coupling reveals that the Brownian bridge is not stationary. But, by truncation this coupling is removed, and the truncated process becomes stationary. \\

\noindent {\bf Acknowledgements.} The authors acknowledge the support of NSF-DMS 1208853 and NSF-DMS 1412343 for this work. \\

\noindent {\bf References:}

\begin{enumerate}

\item Aronszajn, N. (1950). Theory of reproducing kernels. {\it Transactions of the American Mathematical Society}, {\bf 68}, 337 - 404.

\item Chil\`es, J., and Delfiner, P. (2012). \emph{Geostatistics: Modeling Spatial Uncertainty}, 2nd Edition, Wiley, New York.

\item Craven, P. and Wahba, G. (1979). Smoothing noisy data with spline functions. {\it Numerische Mathematik}. {\bf 31,} 377-403.


\item Cressie, N. (1989). Geostatistics. \emph{The American
  Statistician}. Vol. 43, 197-202.

\item Cressie, N. (1990). Reply to ``Comment on Cressie'' by
G. Wahba. \emph{The American Statistician}. Vol. 44, 256-258.

\item \noindent Cressie, N. (1993). {\em Statistics for Spatial Data}, revised ed. Wiley, New York.

\item Dufour, J.-M. and Roy, R. (1976). On spectral estimation for a homogeneous random process on the circle. {\it Stochastic Processes and their Applications.} {\bf 4}, 107-120.

\item Eubank, R.L. (1988). {\it Spline smoothing and nonparametric regression.} Marcel-Dekker, New York, NY.

\item Furrer, E. M. and Nychka, D. W. (2007). A framework to understand the asymptotic properties of kriging and splines. {\it Journal of the Korean Statistical Society.} {\bf 36,} 57-76.


\item Gneiting, T. (1998). Simple test for the validity of correlation function models on the circle. {\it Statistics and Probability Letters},  {\bf 39}, 119-122.

\item Huang, C., Zhang, H., and Robeson, S. (2011). On the validity of commonly used covariance and variogram functions on the sphere. {\it Mathematical Geosciences}, {\bf 43}, 721 - 733.

\item Kent, J. T. and Mardia, K. V. (1994). The link between kriging and thin-plate splines. {\it Probability, Statistics, and Optimization: A tribute to Peter Whittle,} ed. F.P.Kelly, Chichester: Wiley, 325-339.

\item Laslett, G.F. (1994). Kriging and splines: An empirical comparison of their predictive performance in some applications. {\it Journal of American Statistics Association}, {\bf 89}, 391-400.

\item Levesley J, Light W., Ragozin, D., and Sun, X. (1999). A simple approach to the variational theory for interpolation on spheres. {\it International Series of Numerical Mathematics}, {\bf 132}, Birkhauser, Switzerland.

\item Light, W.A. and Wayne, H.S.J. (1995). Error estimates for approximation by radial basis functions. {\it Approximation Theory, Wavelets and Applications}. ed. S.P. Singh, Kluwer Academic, Dordrecht, 215-246.

\item Lorenc, A. C. (1986). Analysis methods for numerical weather
prediction. \emph{Quarterly Journal of the Royal Meteorological
  Soceity}. Vol. 112, 1177-1194.

\item Matheron, G. (1973). The intrinsic random functions and their applications. {\em Advances in Applied Probability}, 5, 439-468.

\item Matheron, G. (1979). Comment translater les catastrophes. La structure des F.A.I. \'{g}en\'{e}rales. Manuscript
N-167, Centre de G\'{e}ostatistique, Fontainebleau, France, 1-36.

\item Matheron, G. (1981). Splines and kriging: their formal equivalence. {\it Syracuse University Geological Contributions}, Syracuse, NY, 77-95.

\item Roy, R. (1972). Spectral analysis for a random process on the circle. {\it Journal of Applied Probability}, {\bf 9}, 745-757.

\item Roy, R., and Dufour, J.-M. (1974). Exact properties of spectral estimates for a Gaussian process on the circle. \emph{Utilitas Mathematica}, 5, 281-291.

\item Schoenberg, I.J. (1942). Positive definite functions on spheres. {\it Duke Mathematics Journal}, {\bf 9}, 96-108.

\item Taijeron, H.J. Gibson, A.G., and Chandler, C. (1994). Spline interpolation and smoothing on hyperspheres. {\it SIAM Journal on Scientific Computing,} {\bf 15}, 1111-1125.

\item Watson, G. S. (1984). Smoothing and interpolation by kriging with splines. \emph{Mathematical Geology}. Vol. 16, 601-615.

\item Wahba, G. (1990a). {\it Spline Models for Observational Data}. { CBMS-NSF regional conference series in applied mathematics}, {\bf 59}, Philadelphia, PA: Society for Industrial and Applied Mathematics.

\item Wahba, G. (1990b). Comment on Cressie. \emph{The American Statistician}. Vol. 44, 255-256.


\item Wood, A. (1995). When is a truncated covariance function on the line a covariance function on the circle? {\it Statistics and Probability Letters}, {\bf 24}, 157-164.

\item Yaglom, A. M. (1961). Second-order homogeneous random fields. {\it Fourth Berkeley Symposium on Mathematical Statistics and Probability}, {\bf 2}, 593 - 622, Berkeley, University of California Press.

\end{enumerate}

\end{document}